\documentclass[11pt]{article}
\pdfoutput=1
\usepackage{amsmath,epsf,amssymb,latexsym,array,color}
\usepackage{graphicx,wasysym}
\usepackage{cancel}
\usepackage{subfig}

\def\be{\begin{equation}}
\def\ee{\end{equation}}
\def\ba{\begin{array}}
\def\ea{\end{array}}
\newcommand{\beq}{\begin{equation}}
\newcommand{\eeq}[1]{\label{#1}\end{equation}}
\newcommand{\bea}{\begin{eqnarray}}
\newcommand{\eea}[1]{\label{#1}\end{eqnarray}}

\newcommand{\bbox}{\lower.2ex\hbox{$\Box$}}
\csname @addtoreset\endcsname{equation}{section}
\usepackage{ifpdf}
\ifx\pdfoutput\undefined
   \pdffalse
   \usepackage{cite}
 \else
   \pdfoutput=1
   \pdftrue
  \usepackage[pdftex]{hyperref}
\def\be{\begin{equation}}
\def\ee{\end{equation}}

\usepackage{ifpdf}
\ifx\pdfoutput\undefined
   \pdffalse
\else
   \pdfoutput=1
   \pdftrue
  \usepackage[pdftex]{hyperref}
\begin{document}
%\title{Constraining Liberated Supergravity}
%\author{Hun Jang}
%\email{hun.jang@nyu.edu}
%\author{Massimo Porrati}
%\email{massimo.porrati@nyu.edu, hun.jang@nyu.edu}
%\affiliation{Center for Cosmology and Particle Physics\\ Department of Physics, New York University \\
%726 Broadway, New York NY 10003, USA}

%\maketitle

\begin{titlepage}

\hskip 1.5cm

\begin{center}
{\huge \bf{Supermassive gauginos in supergravity inflation with high-scale SUSY breaking}}
\vskip 0.8cm  
{\bf \large Hun Jang$^{\star\dag\sharp}$\footnote{hun.jang@nyu.edu} and Massimo Porrati$^{\ddag}$\footnote{massimo.porrati@nyu.edu}}  
\vskip 0.75cm
{\em $^{\star}$Center for Quantum Spacetime (CQUeST), Sogang University, \\ 35 Baekbeom-ro, Mapo-gu, Seoul 04107, Republic of Korea, \\ $^{\dag}$Yukawa Institute for Theoretical Physics (YITP), Kyoto University,\\
Kitashirakawa Oiwakecho, Sakyo-ku, Kyoto 606-8502, Japan and\\ $^{\sharp}$Research Institute of Basic Sciences (RIBS), Incheon National University, \\ 119, Academy-ro, Yeonsu-gu, Incheon 22012, Republic of Korea\\ 
    $^{\ddag}$Center for Cosmology and Particle Physics\\
	Department of Physics, New York University\\
	726 Broadway, New York, NY 10003, USA}
\vspace{12pt}

\end{center}

\begin{abstract}
  A model of supergravity inflation we recently proposed can produce slow roll
  inflation and a realistic spectrum of particles even without F-term supersymmetry breaking. Supersymmetry is broken only by a D-term induced by a recently discovered new type of Fayet-Iliopoulos (FI) term.  Almost
  all supersymmetric partners of
  the standard model fields can get masses as high as the inflationary Hubble scale. The exception is gauginos, for which the vanishing of F-terms implies an
  exact
  cancellation that keeps their masses exactly zero.
  To cure this problem without
  spoiling the simplicity of our model we introduce a new term that further
  enlarges the space of supergravity effective actions. It is an F-term that,
  similarly to the  new FI term,
  becomes singular in the supersymmetric limit. We show that this term can produce  large gaugino masses without altering the spectrum of other states and without lowering the cutoff of the effecive theory. 

\end{abstract}

\vskip 1 cm
\vspace{24pt}
\end{titlepage}
\tableofcontents

\section{Introduction}

 In standard supergravity actions gauginos are massive only if the
  F and/or D term of matter fields are non-vanishing~\cite{fvp}. This requires
  nonvanishing vacuum expectation values (VEV)s of matter fields which may
  generate theoretical difficulty. First, they can make a theory less predictive because it often needs many additional scalars (moduli). Second, the scalar potential necessary to have both nonvanishing F and D terms and stable moduli  is
  neither simple nor natural. Finally, the scalar potential required by moduli stabilization and realistic gaugino masses may be incompatible with  the scalar field dynamics that we wish to study within the theory. 

  On the other hand, models with vanishing VEVs of matter fields carry their own
  set of problems. A serious one is that the lack of experimental evidence for superpartners \cite{Workman:2022ynf} requires among other things heavy gauginos,
  which are difficult to obtain in supergravity models where the VEVs of
  matter fields vanish. It is thus natural to ask the question: {\it can we obtain large gaugino masses together with vanishing VEVs of matter fields and a
    simple moduli stabilization mechanism?} In this paper we will
  answer in the affirmative by explicitly presenting a model with all the
  required properties. 

  An action containing the so-called new Fayet-Iliopoulos (FI) terms \cite{cftp,acik}, which is specified by a hidden-sector vector multiplet $V$, does not contribute to the masses of matter fermions and gauginos, except for the gaugino of the new-FI vector $V$, but it can make matter scalars (or equivalently sfermions) parametrically heavier than observable-sector states. While the new FI terms
  do not give mass to gauginos, other terms can be added to the standard supergravity action that do it. In this paper we will describe in details in
  Section 2 one such term: a new F-term that couples the gauge vector multiplets to $V$. 

  Before embarking in a description of our model, we need to mention several
  works
  relevant to the problem of gaugino masses.  First, non-universal gaugino masses can be obtained in non-minimal GUT models \cite{Ellis:1984bm,Ellis:1985jn,Drees:1985bx,Anderson:1996bg,Martin:2009ad,Younkin:2012ui,Kobayashi:2017fgl},  mirage mediation \cite{Choi:2004sx,Choi:2005ge,Endo:2005uy,Choi:2005uz,Choi:2005hd,Kitano:2005wc,Choi:2006xb}, and non-universal gauge kinetic functions through string compactifications \cite{Blumenhagen:2006ci}. Other recent developments in the non-universal gaugino masses were discussed in~\cite{Chakrabortty:2013voa,Martin:2013aha,Chakrabortty:2015ika,Sumita:2015tba, Kawamura:2018qda}. Second, gaugino masses can be obtained by anomaly mediation \cite{Randall:1998uk, Giudice:1998xp, Harigaya:2015xia} in the heavy sfermion scenario, which are proportional to the inverse of the corresponding gauge coupling constant $g$, its beta function $\beta(g^2)$, and gravitino mass $m_{3/2}$. In addition, anomaly-mediated gaugino mass can be derived in the superspace formulation of supergravity with a Wilsonian effective action \cite{Harigaya:2014sfa}, using superconformal anomaly \cite{Pomarol:1999ie}, and using the anomaly of the Peccei-Quinn (PQ) symmetry mediated by additional charged matters under SM gauge symmetries \cite{Peccei:1977hh,Peccei:1977ur,Weinberg:1977ma,Wilczek:1977pj} for KSVZ-type QCD axion \cite{Kim:1979if,Shifman:1979if}. Furthermore, gaugino masses get corrections as large as the anomalous one if vector-like matter fields with masses smaller than the gravitino one exist \cite{Nelson:2002sa,Hsieh:2006ig, Gupta:2012gu}. Let us point out that the common point of all those  models for gaugino masses is that moduli stabilization with {\it non-vanishing VEVs of matter fields} is necessary and deeply associated with generating the gaugino masses, signaling a possible backreaction between the dynamics of scalar fields and gaugino masses. On the contrary, our model for the gaugino-mass generation can decouple moduli stabilization from scalar field dynamics while keeping {\it vanishing VEVs of all matter fields} (except the Higgs fields).

This paper describes a mechanism for generating arbitrarily large of gaugino masses in a simple extension of the
 supergravity model of inflation
with new Fayet-Iliopoulos (FI) terms~\cite{cftp,acik} presented in~\cite{JP4}. In~\cite{JP4} we used the K\"ahler form of the 
new FI term given in~\cite{acik} to construct a model with slow roll inflation, and realistic spectrum and interactions. 
Among other features, our 
model contained the complete spectrum of the minimal supersymmetric standard model (MSSM) together with its 
interactions and terms necessary to give large masses to {\em almost all}  the supersymmetric partners of standard model (SM) particles. The exception was gaugino masses. Eq.~(77) of Ref.~\cite{JP4} shows that they are proportional to the F term $WG_J=K_JW+W_J$, where $K$ is the K\"ahler potential, $W$ is the superpotential and the indices $J$ are
either those of the modulus $T$ or those of the matter scalars $z^i$. Ref.~\cite{JP4} overlooked the simple fact that the 
sum $K_JW+W_J$ is precisely an F term so incorrectly assumed that it was of the same order of its two summands $K_JW$ and $W_J$. So it stated that gaugino masses can be made $\mathcal{O}(H)$, with $H$, the Hubble parameter
during slow-roll inflation. The gaugino masses are instead zero unless there is an F-term supersymmetry breaking. 
A nonzero F-term would spoil the most appealing feature of the model; namely that a parametrically
large mass splitting between standard model fermions and supersymmetric scalar partners can be achieved even with a 
pure D-term supersymmetry breaking, where {\em all}  F terms vanish at the post-inflationary minimum ($T=1/2,z^i\sim 0$).
It would also require significantly more convoluted superpotentials than those given in~\cite{JP4}, which
 could spoil good properties of the model such as the possibility of slow roll inflation, and in 
 any case would only produce gaugino masses much smaller than those of the other superpartners. The solution we present here is different: we add a new gaugino mass (GM) term that, much like the new FI terms, 
 becomes singular when supersymmetry is unbroken. It is an F term of conformal weight 3 whose schematic form in the standard superfield notations is 
 \beq
 \int d^2 \theta W_\alpha (U) W^\alpha (U) W_\beta(V) W^\beta(V) S_0^{-3} W(z,T)^{-1} ,
 \eeq{intro1}
 where $V$ is the $U(1)$ vector superfield associated to the new FI term of Ref.~\cite{acik}, $U$ are the vector superfields
 of the observable sector, $W$ is as before the superpotential and $S_0$ is the conformal compensator. The detailed
  construction of  the GM term using superconformal tensor calculus is given in Sec. 2. The explicit form of the gaugino masses and an analysis of the range of masses allowed by the GM
  term, together with their relation to the cutoff and to the Hubble scale in inflation is given in Sec. 3. 
 Section 4 contains a detailed analysis of the effect of the GM term on the cutoff of the 
 effective field theory. Since this new term contains inverse powers of F terms, it could in principle lower the cutoff
 below the inflationary Hubble scale $H$. This would make the effective theory useless to describe 
 inflation. Section 3 finds the range of the free  parameters contained in the GM term that keeps the effective 
 theory cutoff $ \Lambda$ larger than $H$.

\section{Gaugino masses in supergravity with new Fayet-Iliopoulos terms}

In this section, we describe a mechanism for generating arbitrarily large of gaugino masses in supergravity with new Fayet-Iliopoulos (FI) terms.  While the gaugino of the vector multiplet for the new FI term gets a heavy mass from the new FI term, the gauginos of the SM gauge vector multiplets will get non-vanishing masses, as heavy as the Hubble scale via the new 
GM F term~\eqref{intro1} .

As in Ref.~\cite{JP4} we work in $M_{pl}=1$ units and introduce matter chiral multiplets $Z^i$, the chiral compensator $S_0$, a real multiplet $V$, and another real multiplet 
$(V)_D$, whose lowest component is the auxiliary D term of the real multiplet $V$. The components of the various superconformal multiplets are given as follows. 
\begin{description}
\item{Chiral matter multiplets}
\begin{eqnarray}
&& Z^i = (z^i,-i\sqrt{2}P_L\chi^i,-2F^i,0,+i\mathcal{D}_{\mu}z^i,0,0) = \{ z^i, P_L\chi^i,F^i\},\\
&& \bar{Z}^{\bar{i}} = (\bar{z}^{\bar{i}},+i\sqrt{2}P_R\chi^{\bar{i}},0,-2\bar{F}^{\bar{i}},-i\mathcal{D}_{\mu}\bar{z}^{\bar{i}},0,0) = \{ \bar{z}^{\bar{i}}, P_R\chi^{\bar{i}},\bar{F}^{\bar{i}}\}.
\end{eqnarray}
\item{Conformal compensator multiplet}
\begin{eqnarray}
&& S_0 = (s_0,-i\sqrt{2}P_L\chi^0,-2F_0,0,+i\mathcal{D}_{\mu}s_0,0,0) = \{ s_0, P_L\chi^0,F_0\},\\
&& \bar{S}_0 = (\bar{s}_0,+i\sqrt{2}P_R\chi^0,0,-2\bar{F}_0,-i\mathcal{D}_{\mu}\bar{s}_0,0,0) = \{\bar{s}_0, P_R\chi^0,\bar{F}_0\}.    
\end{eqnarray}
\item{Field strength multiplet for the new FI vector $V$}
\begin{eqnarray}
V &=& \{0,0,0,0,v_{\mu},\lambda_V,D_V\} ~\textrm{in the Wess-Zumino gauge}\\
(\bar{\lambda}P_L\lambda)_V &=& (\bar{\lambda}_VP_L\lambda_V,-i\sqrt{2}P_L\Lambda_V,2D_-^2,0,+i\mathcal{D}_{\mu}(\bar{\lambda}_VP_L\lambda_V),0,0) \nonumber\\
&=& 
\{\bar{\lambda}_VP_L\lambda_V, P_L\Lambda_V,-D_-^2\},\label{FSV_multiplet}\\
 (\bar{\lambda}P_R\lambda)_V &=& (\bar{\lambda}_VP_R\lambda_V,+i\sqrt{2}P_R\Lambda_V,0,2D_+^2,-i\mathcal{D}_{\mu}(\bar{\lambda}_VP_R\lambda_V),0,0) \nonumber\\
 &=& \{\bar{\lambda}_VP_R\lambda_V, P_R\Lambda_V,-D_+^2\},
\end{eqnarray}
where
\begin{eqnarray}
&& P_L\Lambda_V \equiv \sqrt{2}P_L(-\frac{1}{2}\gamma\cdot \hat{F}_V + iD_V)\lambda_V,\qquad P_R\Lambda_V \equiv \sqrt{2}P_R(-\frac{1}{2}\gamma\cdot \hat{F}_V - iD_V)\lambda_V,\\
&& D_{V-}^2 \equiv D_V^2 - \hat{F}_V^-\cdot\hat{F}_V^- - 2  \bar{\lambda}_VP_L\cancel{\mathcal{D}}\lambda_V,\qquad D_{V+}^2 \equiv D_V^2 - \hat{F}_V^+\cdot\hat{F}_V^+ - 2  \bar{\lambda}_VP_R\cancel{\mathcal{D}}\lambda_V,\\
&& \mathcal{D}_{\mu}\lambda_V \equiv \bigg(\partial_{\mu}-\frac{3}{2}b_{\mu}+\frac{1}{4}w_{\mu}^{ab}\gamma_{ab}-\frac{3}{2}i\gamma_*\mathcal{A}_{\mu}\bigg)\lambda_V - \bigg(\frac{1}{4}\gamma^{ab}\hat{F}_{Vab}+\frac{1}{2}i\gamma_* D_V\bigg)\psi_{\mu}
\\
 && \hat{F}_{Vab} \equiv F_{Vab} + e_a^{~\mu}e_b^{~\nu} \bar{\psi}_{[\mu}\gamma_{\nu]}\lambda_V,\qquad F_{Vab} \equiv e_a^{~\mu}e_b^{~\nu} (2\partial_{[\mu}v_{\nu]}),\\
 && \hat{F}^{\pm}_{V\mu\nu} \equiv \frac{1}{2}(\hat{F}_{V\mu\nu}\pm \tilde{\hat{F}}_{V\mu\nu}), \qquad \tilde{\hat{F}}_{V\mu\nu} \equiv -\frac{1}{2} i\epsilon_{\mu\nu\rho\sigma}\hat{F}_V^{\rho\sigma} .
\end{eqnarray}
\item{Field strength multiplet for the SM gauge vector $U$}
\begin{eqnarray}
U &=& \{0,0,0,0,u_{\mu},\lambda_U,D_U\} ~\textrm{in the Wess-Zumino gauge}\\
 (\bar{\lambda}P_L\lambda)_U &=& (\bar{\lambda}_U P_L\lambda_U,-i\sqrt{2}P_L\Lambda_U,2D_{U-}^2,0,+i\mathcal{D}_{\mu}(\bar{\lambda}_U P_L\lambda_U),0,0) \nonumber\\
 &=& \{\bar{\lambda}_U P_L\lambda_U, P_L\Lambda_U,-D_{U-}^2\}\label{FSU_multiplet},\\
 (\bar{\lambda}P_R\lambda)_U &=& (\bar{\lambda}_U P_R\lambda_U,+i\sqrt{2}P_R\Lambda_U,0,2D_{U+}^2,-i\mathcal{D}_{\mu}(\bar{\lambda}_U P_R\lambda_U),0,0)\nonumber\\
 &=& \{\bar{\lambda}_U P_R\lambda_U, P_R\Lambda_U,-D_{U+}^2\},
\end{eqnarray}
where
\begin{eqnarray}
&& P_L\Lambda_U \equiv \sqrt{2}P_L(-\frac{1}{2}\gamma\cdot \hat{F}_U + iD_U)\lambda_U,\qquad P_R\Lambda_U \equiv \sqrt{2}P_R(-\frac{1}{2}\gamma\cdot \hat{F}_U - iD_U)\lambda_U,\\
&& D_{U-}^2 \equiv D_U^2 - \hat{F}_U^-\cdot\hat{F}_U^- - 2  \bar{\lambda}_UP_L\cancel{\mathcal{D}}\lambda_U,\qquad D_{U+}^2 \equiv D_U^2 - \hat{F}_U^+\cdot\hat{F}_U^+ - 2  \bar{\lambda}_UP_R\cancel{\mathcal{D}}\lambda_U,\\
&& \mathcal{D}_{\mu}\lambda_U \equiv \bigg(\partial_{\mu}-\frac{3}{2}b_{\mu}+\frac{1}{4}w_{\mu}^{ab}\gamma_{ab}-\frac{3}{2}i\gamma_*\mathcal{A}_{\mu}\bigg)\lambda_U - \bigg(\frac{1}{4}\gamma^{ab}\hat{F}_{Uab}+\frac{1}{2}i\gamma_* D_U\bigg)\psi_{\mu}
\\
 && \hat{F}_{Uab} \equiv F_{Uab} + e_a^{~\mu}e_b^{~\nu} \bar{\psi}_{[\mu}\gamma_{\nu]}\lambda_U,\qquad F_{Uab} \equiv e_a^{~\mu}e_b^{~\nu} (2\partial_{[\mu}u_{\nu]}),\\
 && \hat{F}^{\pm}_{U\mu\nu} \equiv \frac{1}{2}(\hat{F}_{U\mu\nu}\pm \tilde{\hat{F}}_{U\mu\nu}), \qquad \tilde{\hat{F}}_{U\mu\nu} \equiv -\frac{1}{2} i\epsilon_{\mu\nu\rho\sigma}\hat{F}_U^{\rho\sigma} .
\end{eqnarray}
\end{description}

The GM F-term Lagrangian, written in the notations of superconformal tensor calculus of~\cite{fvp}, is
\begin{eqnarray}
\mathcal{L}_{GM} =[(\bar{\lambda}P_L\lambda)_U (\bar{\lambda}P_L\lambda)_V S_0^{-3} W(Z)^{-1}]_F + c.c.,\label{GM}
\end{eqnarray}
where $W(Z)$ is the superpotential of our theory, and $(\bar{\lambda}P_L\lambda)_U$ and $(\bar{\lambda}P_L\lambda)_V$ are the composite multiplets corresponding to each of the two vectors. It is an F-term density of conformal weight 3
and chiral weight 3, so it is invariant under supersymmetry. 
The superpotential multiplet is given by
\begin{eqnarray}
    W(Z) = \Big\{W(z), W_iP_L\chi^i, W_iF^i - \frac{1}{2} W_{ij}\bar{\chi}^iP_L\chi^j\Big\}.
\end{eqnarray}
To compute the Lagrangian \eqref{GM}, we can use the following results of superconformal tensor calculus
\begin{eqnarray}
 W^{-1}(Z) &=& \Big\{\frac{1}{W}, -\frac{W_i}{W^2}P_L\chi^i, -\frac{W_iF^i}{W^2}+\Big(\frac{W_{ij}}{2W^2}+\frac{W_iW_j}{W^3}\Big)\bar{\chi}^iP_L\chi^j\Big\},\label{WInv_multiplet}\\
 S_0^{-3} &=& \Big\{ 
 \frac{1}{s_0^3}, -\frac{3}{s_0^4}P_L\chi^0, -\frac{3F^0}{s_0^4}
 \Big\}.\label{Sinv3_multiplet}
\end{eqnarray}
Multiplying the supermultiplets we get
\begin{eqnarray}
&&S_0^{-3}W^{-1}(Z) \nonumber\\
&&= \Big\{ \frac{1}{s_0^3W},
 -\frac{3}{s_0^4W}P_L\chi^0  -\frac{W_i}{s_0^3W^2}P_L\chi^i, \nonumber\\&&
\qquad -\Big(\frac{3F^0}{s_0^4W} +\frac{W_iF^i}{s_0^3W^2}\Big)+\Big(\frac{W_{ij}}{2s_0^3W^2}+\frac{W_iW_j}{s_0^3W^3}\Big)\bar{\chi}^iP_L\chi^j
 - \frac{3W_i}{s_0^4W^2} \bar{\chi}^i P_L \chi^0
 \Big\}. \label{multi1}
\end{eqnarray}
The multiplication of the field strength mutliplets of $V$ and $U$ gives
\begin{eqnarray}
 && (\bar{\lambda}P_L\lambda)_U (\bar{\lambda}P_L\lambda)_V  \nonumber\\
 &&=\{ (\bar{\lambda}_U P_L\lambda_U) (\bar{\lambda}_V P_L\lambda_V), \nonumber\\
 &&\qquad 
(\bar{\lambda}_U P_L\lambda_U) \sqrt{2}P_L(-\frac{1}{2}\gamma\cdot \hat{F}_V + iD_V)\lambda_V + (\bar{\lambda}_V P_L\lambda_V)\sqrt{2}P_L(-\frac{1}{2}\gamma\cdot \hat{F}_U + iD_U)\lambda_U, \nonumber\\
&& \qquad 
 -(\bar{\lambda}_U P_L\lambda_U)(D_V^2 - \hat{F}_V^-\cdot\hat{F}_V^- - 2  \bar{\lambda}_VP_L\cancel{\mathcal{D}}\lambda_V)-(\bar{\lambda}_V P_L\lambda_V)(D_U^2 - \hat{F}_U^-\cdot\hat{F}_U^- - 2  \bar{\lambda}_UP_L\cancel{\mathcal{D}}\lambda_U)
 \nonumber\\
 &&\qquad -2\bar{\lambda}_U(-\frac{1}{2}\gamma\cdot \hat{F}_U + iD_U)  P_L(-\frac{1}{2}\gamma\cdot \hat{F}_V + iD_V)\lambda_V
 \}. \label{multi2}
\end{eqnarray}
Since the Lagrangian~\eqref{GM} is the F term of the two composite multiplets~\eqref{multi1} and~\eqref{multi2}, we find 
\begin{eqnarray}
\mathcal{L}_{\textrm{GM}}
 &=& \bigg[ -\frac{1}{s_0^3W}(\bar{\lambda}_U P_L\lambda_U)(D_V^2 - \hat{F}_V^-\cdot\hat{F}_V^- - 2  \bar{\lambda}_VP_L\cancel{\mathcal{D}}\lambda_V)
 \nonumber\\
 &&-\frac{1}{s_0^3W}(\bar{\lambda}_V P_L\lambda_V)(D_U^2 - \hat{F}_U^-\cdot\hat{F}_U^- - 2  \bar{\lambda}_UP_L\cancel{\mathcal{D}}\lambda_U)
 \nonumber\\
 && -2\frac{1}{s_0^3W}\bar{\lambda}_U(-\frac{1}{2}\gamma\cdot \hat{F}_U + iD_U)  P_L(-\frac{1}{2}\gamma\cdot \hat{F}_V + iD_V)\lambda_V
 \nonumber\\
 &&  -\Big(\frac{3F^0}{s_0^4W} +\frac{W_iF^i}{s_0^3W^2}\Big)(\bar{\lambda}_U P_L\lambda_U) (\bar{\lambda}_V P_L\lambda_V)\nonumber\\
 && +\Big(\frac{W_{ij}}{2s_0^3W^2}+\frac{W_iW_j}{s_0^3W^3}\Big)\bar{\chi}^iP_L\chi^j(\bar{\lambda}_U P_L\lambda_U) (\bar{\lambda}_V P_L\lambda_V)
\nonumber\\
&&- \frac{3W_i}{s_0^4W^2} (\bar{\chi}^i P_L \chi^0)(\bar{\lambda}_U P_L\lambda_U) (\bar{\lambda}_V P_L\lambda_V) \nonumber\\
&&+ \Big(\frac{3}{s_0^4W}\bar{\chi}^0P_L  + \frac{W_i}{s_0^3W^2}\bar{\chi}^iP_L\Big)(\bar{\lambda}_U P_L\lambda_U) \sqrt{2}P_L(-\frac{1}{2}\gamma\cdot \hat{F}_V + iD_V)\lambda_V
\nonumber\\
&& +\Big(\frac{3}{s_0^4W}\bar{\chi}^0P_L  + \frac{W_i}{s_0^3W^2}\bar{\chi}^iP_L\Big) (\bar{\lambda}_V P_L\lambda_V)\sqrt{2}P_L(-\frac{1}{2}\gamma\cdot \hat{F}_U + iD_U)\lambda_U \bigg] + c.c. \label{gm}
\end{eqnarray}
The GM term produces another contribution to the solutions of the equations of motion for the auxiliary field of the compensator $S_0$, $F^0$, and for the compensators $F^j$ of the physical chiral multiplets. We write it inside large brackets
\begin{eqnarray}
 F^0 &=& e^{2K/3} \bar{W}(\bar{z}) - \frac{1}{3} e^{K/6} (\bar{\lambda}P_R\lambda)_V +  \Biggl[ s_0^{*-4}\Bar{W}^{-1}
 (\bar{z})e^{K/3} (\bar{\lambda}P_R\lambda)_U(\bar{\lambda}P_R\lambda)_V \Biggr]  , \nonumber \\ 
  F^i &=& -3e^{K/2}G^{i\bar{j}}(\bar{W}_{\bar{j}}+K_{\bar{j}}\bar{W}) -G^{i\bar{j}}\Big(9 \frac{\mathcal{U}_{\bar{j}}}{\mathcal{U}}+ 3K_{\bar{j}}\Big)(\bar{\lambda}P_R\lambda)_V \nonumber \\
& &+ \Biggl[  9 s_0^{*-3}\bar{W}_{\bar{j}}(\bar{z})\bar{W}^{-2}(\bar{z}) (\bar{\lambda}P_R\lambda)_U(\bar{\lambda} 
P_R\lambda)_V  \Biggr] . \label{matter_aux}
\end{eqnarray}
Here $K$ is the K\"{a}hler potential and the second term in Eq.~\eqref{matter_aux} comes from the new FI terms introduced
 in Ref.~\cite{JP4}. Notice that the fermionic component of the composite multiplet defining the Lagrangian~\eqref{GM} contains couplings with three or five fermions only,  so the GM term~\eqref{GM} does not alter the original goldstino contribution to the fermion masses when the gravitino becomes massive. In other words, the GM term contribution to the goldstino mass term produces only higher-order interactions, not mass terms. Therefore, all the mass matrices except  those of the gaugini remain the same as those of Ref.~\cite{JP4}.

\section{Supermassive gauginos in supergravity inflation with high-scale SUSY breaking}
The contribution of~\eqref{gm} to gaugino masses is contained in the terms
\begin{eqnarray}
    \mathcal{L}_{\textrm{2 fermions}} &\supset&  (\bar{\lambda}P_L\lambda)_V s_0^{-3} W(z)^{-1}(-D_{U-}^2) + (\bar{\lambda}P_L\lambda)_U s_0^{-3} W(z)^{-1} (-D_{V-}^2) \nonumber\\
    && +2D_U D_V s_0^{-3} W(z)^{-1} \bar{\lambda}_UP_L\lambda_V+2D_U D_V s_0^{-3} W(z)^{-1} \bar{\lambda}_V P_L\lambda_U \label{boson_GM}
\end{eqnarray}
where $z,s_0$ are complex scalars, and $D_{U/V-}^2 \equiv D_{U/V}^2 - \hat{F}_{U/V}^{-}\cdot\hat{F}_{U/V}^{-} - 2  (\bar{\lambda}P_L\cancel{\mathcal{D}}\lambda)_{U/V}$. Notice that there are off diagonal quadratic terms in the gaugino fermions. The other terms in~\eqref{gm} merely produce matter-fermion couplings and gaugino interactions. In general, since the Lagrangian \eqref{boson_GM} contains off diagonal quadratic terms, we need to diagonalize the gaugino mass matrix 
\begin{eqnarray}
M_{\textrm{Gaugino}} =  \frac{1}{W(z)s_0^3} 
\begin{pmatrix}
-D_U^2 & 2D_UD_V \\
2D_VD_U & - D_V^2.
\end{pmatrix}
\end{eqnarray}
However, the D terms of the SM vector multiplet for $SU(3)_c$ vanish, while those for $U(1)_Y$ and $SU(2)_L$ are given by $D_U \sim \mathcal{O}(g_a(v_u^2-v_d^2))$, where $g_a$'s are the SM ($U(1)_Y,SU(2)_L$) gauge coupling constant and $v_{u,d}$ are the vacuum expectation values of the two Higgs fields, $H_u$ and $H_d$, present in the MSSM. 
These
D terms are much smaller than those of the new FI term vector $V$; therefore, we can neglect them and write the Lagrangian as follows
\begin{eqnarray}
      &&  \mathcal{L}_{\textrm{Gaugino-mass}} \supset -\frac{D_{V}^2}{W(z)s_0^3} (\bar{\lambda}P_L\lambda)_U = -\frac{D_{V}^2}{W(z)e^{K/2}} (\bar{\lambda}P_L\lambda)_U = -\frac{D_V^2}{m_{3/2}}(\bar{\lambda}P_L\lambda)_U \nonumber\\
      &&  \implies M_{\textrm{gaugino}} = \frac{D_V^2}{m_{3/2}} \sim H.
        \label{boson_GM2}
\end{eqnarray}
Here we used the superconformal gauge-fixing condition on the compensator $s_0 = e^{K/6}$, and the gravitino mass relation $m_{3/2} = W(z)e^{K/2}$. Notice that the gaugino masses of the three SM gauge groups can be as large as
 the inflationary Hubble scale $H$, both during and after inflation, as long as the gravitino mass remains of 
 $\mathcal{O}(H)$ during and after slow-roll inflation. 
\begin{eqnarray}
 D_V \sim \xi = M_S^2 \sim H, \qquad m_{3/2} \sim H. 
\end{eqnarray}
Here $\xi$ is the mass scale of the inflationary potential and $M_S$ is the supersymmetry breaking scale, both given in  
$M_{pl}=1$ units.

More precisely, the gravitino mass $m_{3/2}$ along the inflationary trajectory is given by
\begin{eqnarray}
    m_{3/2} &=& We^{K/2} = \frac{A(e^{-aT}-c)}{(T+\bar{T})^{3/2}}= \frac{A(e^{-aX/2}-c)}{X^{3/2}}  \nonumber\\
    &=& X^{-3/2} \sqrt{3} H (a^{-1}e^{-a(X-1)/2}-a^{-1}-1/3) ,
\end{eqnarray}
where we use the same superpotential $W = A(e^{-aT}-c)$ as in Ref.~\cite{JP4}, $K = -3\ln[T+\bar{T}]$,  $X=T+\bar{T}$, $T$ is real along the inflationary trajectory where {\it VEVs of the matter scalars are vanishing}, and 
\begin{eqnarray}
    A = a^{-1}e^{a/2}\sqrt{3(M_I^4-\Lambda)}\approx a^{-1}e^{a/2}\sqrt{3} H, \quad c = (1+a/3)e^{-a/2}, \quad M_I^4 \approx H^2 ,
\end{eqnarray}
with $a$ a so-far-unspecified real-number parameter. Our effective theory describes physics only up to a finite cutoff in energy, $\Lambda_{cut}$. The gravitino can be described
by this effective theory only if its mass is below the cutoff (and large enough to be unobservable). To satisfy both constraints
we restrict the range of the gravitino mass to
\begin{eqnarray}
  \alpha H \lesssim  |m_{3/2}| < \Lambda_{cut}, \label{m32}
\end{eqnarray}
where $0 < \alpha < 1$ is another yet-to-be-constrained real parameter. Extra scalar degrees of freedom are required to be heavier than the Hubble scale in order to be integrated out in our single-field inflationary model while fermionic degrees of freedom do not have to be heavier than the Hubble scale, but they still have to but lighter than the cutoff.

{For a very small number $a$, we have
\begin{eqnarray}
    m_{3/2} &\approx& X^{-3/2} \sqrt{3} H \Big(a^{-1}(1-a(X-1)/2)-a^{-1}-1/3\Big) \nonumber\\
    &=& -X^{-3/2} \sqrt{3} H\Big( \frac{X-1}{2}+\frac{1}{3} \Big) = X^{-3/2} \sqrt{3} H \frac{(1-3X)}{6} ,
\end{eqnarray}
so that the allowed range is
\begin{eqnarray}
  \alpha H \lesssim   \frac{\sqrt{3}}{6}X^{-3/2}|1-3X|H  < \Lambda_{cut}.
\end{eqnarray} 
The inflationary trajectory starts from around $X \sim 100$ at the beginning of the slow roll epoch and
ends at the vacuum $X=1$, where the inflaton gets frozen. We are thus interested in the range $1/3 < X$. Then, 
constraint~\eqref{m32} is
\begin{eqnarray}
  \alpha H \lesssim   \frac{\sqrt{3}}{6}X^{-3/2}(3X-1)H  < \Lambda_{cut}.\label{ineq}
\end{eqnarray} 
To investigate this constraint, we define a function 
\begin{eqnarray}
 F(X,\alpha) \equiv  \frac{\sqrt{3}}{6}X^{-3/2}(3X-1) - \alpha,
\end{eqnarray} 
leading to
\begin{eqnarray}
    0 \lesssim F(X,\alpha) < \frac{\Lambda_{cut}}{H} - \alpha .
\end{eqnarray}
We therefore obtain
\begin{eqnarray}
   X &\lesssim& \frac{1}{4 \alpha^2} 
   + \frac{3 -8\alpha^2}{4\alpha^2  \sqrt[3]{3}  (9 - 36 \alpha^2 + 24 \alpha^4 + 8 \alpha^3 \sqrt{-3 + 9 \alpha^2})^{1/3}} \nonumber\\&&+ \frac{(9 - 36 \alpha^2 + 24 \alpha^4 + 8 \alpha^3\sqrt{-3 + 9 \alpha^2})^{1/3}}{4\alpha^2  \sqrt[3]{9}}  ,\label{ineq2}
\end{eqnarray}
which we computed using \url{Mathematica}. To ensure a sufficient number of e-foldings in the slow-roll inflation epoch, 
we set the right side of Eq.~\eqref{ineq2} at $X= 100$. This leads to
\begin{eqnarray}
\alpha \approx 0.086,
\end{eqnarray}
which can also be obtained numerically using \url{Mathematica}. In Fig.~\ref{fig1} we plot $F(X,\alpha)$ vs $X$ for various 
values of $\alpha$ to illustrate this bound.

\begin{figure}[t!]
    \centering
    \includegraphics[width=13cm]{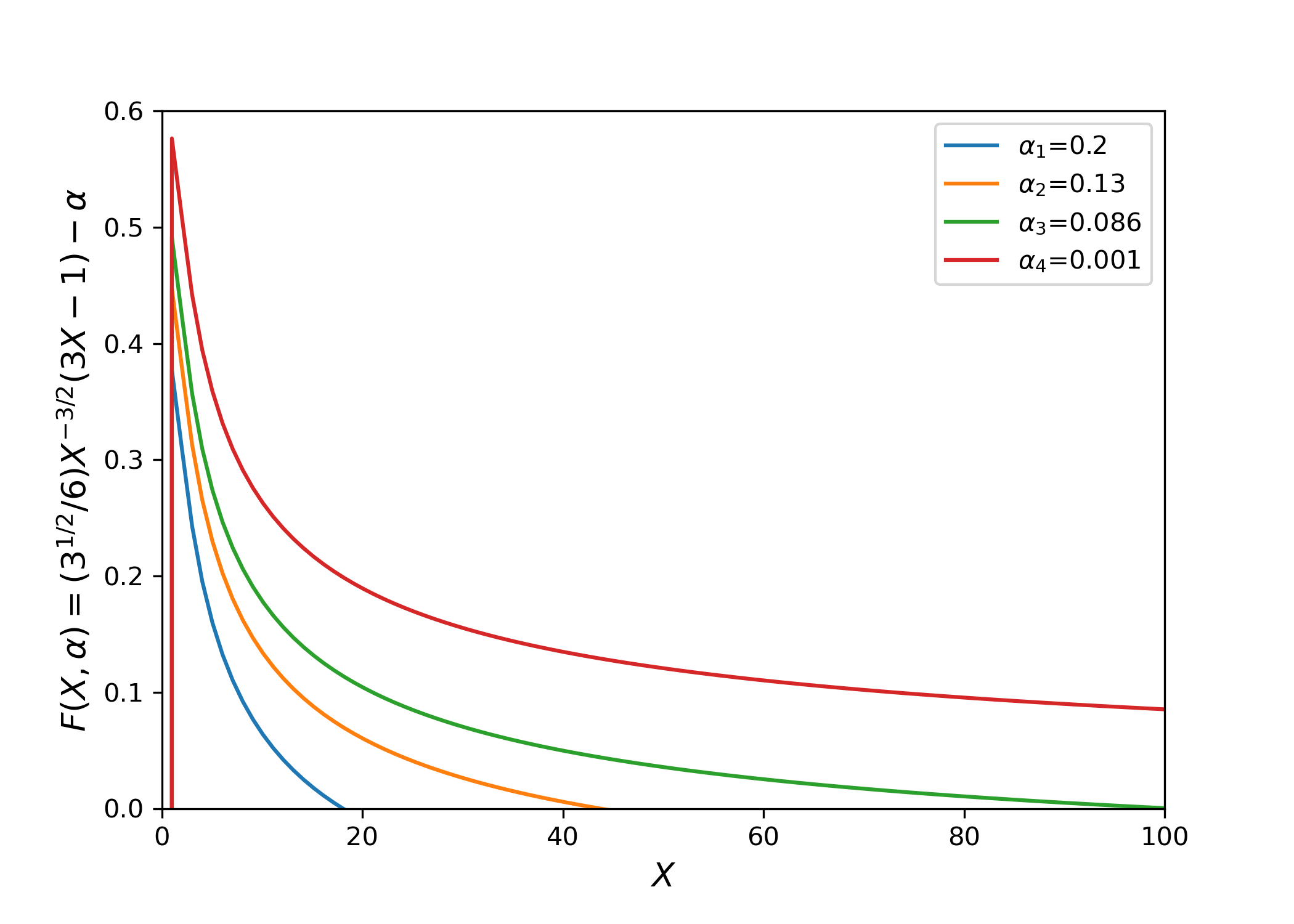}
    \caption{The behaviors of $F(X,\alpha)$ in the direction of $X$ for different $\alpha$'s. We can see that the maximum value of $X$ decreases as $\alpha$ increases. In order to have around 100 e-foldings during slow-roll inflation, 
     $\alpha\sim 0.086$.}
    \label{fig1}
\end{figure}
With this value of $\alpha$, the gravitino mass can remain around the Hubble scale over the whole inflationary trajectory. Re-introducing the reduced Planck mass $M_{pl} ~(\sim 2.4\times 10^{18}~\textrm{GeV)}$ and setting 
$H\sim 10^{-5} M_{pl}$, the gravitino mass bound is given by
\begin{eqnarray}
    0.086 H = 8.6 \times 10^{-7} M_{pl} \lesssim m_{3/2} < \Lambda_{cut} \sim \sqrt{H M_{pl}} \sim 10^{-2.5}M_{pl}
\end{eqnarray}
or equivalently in GeV
\begin{eqnarray}
    10^{12}~\textrm{GeV}  \lesssim m_{3/2} < 10^{15.5}~\textrm{GeV}.
\end{eqnarray}
Since the gaugino mass is given by $M_{Gaugino} \sim \frac{H^2}{m_{3/2}}$, we find that
\begin{eqnarray}
    10^{12}~\textrm{GeV}  \lesssim \frac{H^2}{M_{Gaugino}} \sim \frac{10^{8}~ \textrm{GeV}^2}{M_{Gaugino}} < 10^{15.5}~\textrm{GeV},
\end{eqnarray}
which immediately produces the final result for the gaugino-mass bound
\begin{eqnarray}
10^{10.5} ~\textrm{GeV} < M_{Gaugino} \lesssim 10^{14}~\textrm{GeV} < \Lambda_{cut} = M_S \sim 10^{15.5} ~\textrm{GeV}.
\end{eqnarray}
Note that the gaugino masses can be very large while still being within the above range.

Next, let us see whether the range we consider is consistent with the inflaton dynamics during slow-roll inflation and
reheating. Our scalar potential has the following ``Starobinsky'' form 
\begin{eqnarray}
   V = H^2  - \frac{1}{X^2}\bigg(-2aA^2e^{-aX}+2acA^2 e^{-aX/2} \bigg)+ \frac{1}{3X} a^2A^2 e^{-aX},
\end{eqnarray}
where $A^2 \approx 3a^{-2}e^a M_S^4 = 3a^{-2}e^a H^2$ and $c=(1+a/3)e^{-a/2}$. The redefinition that makes the kinetic term of the inflaton canonically normalized is $X=e^{\sqrt{2/3}\phi}$.  We re-write $V$ 
\begin{eqnarray}
      V= H^2\bigg( 1 - \frac{1}{X^2}\bigg(-2ae^{-aX}+2a(1+a/3)e^{-a/2} e^{-aX/2} \bigg)3a^{-2}e^a+ \frac{1}{X} e^{-a(X-1)}\bigg) \label{Inflaton_potential}
\end{eqnarray}
and note that its minimum is located at $X=1$. For the scalar potential at large $X$ during inflation we have $V \sim H^2M_{pl}^2$. On the contrary, around the point $X=1/3$ (which is NOT a minimum), we have 
\begin{eqnarray}
 V = H^2\bigg(1+ \frac{ - 54 e^{a/3} - 18 a e^{a/3} + 54 e^{2 a/3} + 3 a e^{2 a/3}}{a}\bigg).
\end{eqnarray}
If we assume that $a \ll 1$, after a short calculation and after expanding the exponential terms up to linear order in $a$
we obtain
\begin{eqnarray}
     V \approx 4H^2(1-a) \sim 4H^2.
\end{eqnarray}
This implies that during the reheating era after inflation, the inflaton can fluctuate around the vacuum at $X=1$ without 
reaching the point $X=1/3$, since $V(X=1/3) \sim 4H^2M_{pl}^2$ is greater than $V(X\gg 1) \sim H^2M_{pl}^2$, which 
is the potential at the beginning of slow-roll inflation (and the Hubble friction {\em dissipates} energy). Thus, it is consistent to restrict the range of the inflaton's motion to $1/3 < X \lesssim 100$ along the inflationary trajectory --whose minimum  at $X=1$, as shown in Fig. \ref{Scalar_potential}. In particular, within this range,  the  upper bound in Eq.~\eqref{ineq} is always satisfied, because we have $H\sim \mathcal{O}(10^{-5})$ while, in $M_{pl}=1$ units, $\Lambda_{cut}\sim\mathcal{O}(10^{-2.5})$. 

\begin{figure}[t!]
    \centering
    \includegraphics[width=13cm]{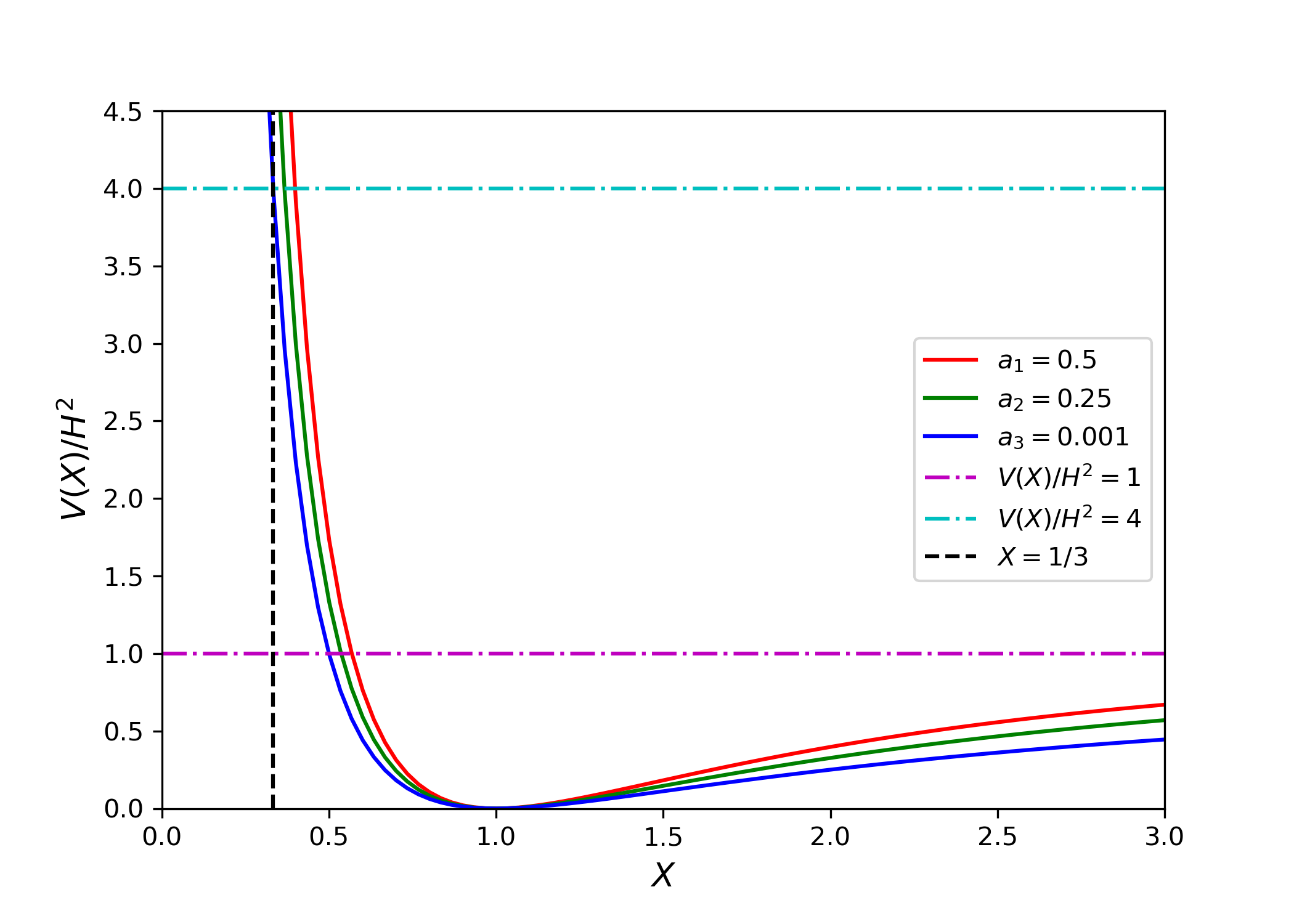}
\caption{The scalar potentials of Starobinsky inflation for different values of $a$ in Eq.~\eqref{Inflaton_potential}, whose minimum is located at $X=1$. Note that during inflation and reheating the potential energy is much smaller than  at $X=1/3$. }    \label{Scalar_potential}
\end{figure}

%\begin{figure}%
%    \centering
%    \subfloat{{\includegraphics[width=8.5cm]{Potential_Big.png} }}%
%    \subfloat{{\includegraphics[width=8.5cm]{Potential_Small.png} }}%
%    \caption{The scalar potentials of Starobinsky inflation for some different values of $a$ in Eq.~\eqref{Inflaton_potential}, whose minimum is located at $X=1$. The left and right plots show the full and short ranges in $X$, respectively. Note that within the range $1/3<X\lesssim 100$, the energy scales along the inflationary trajectory and during the reheating process after inflation are consistently smaller than that of $X=1/3$. {\color{blue} (The figure you would like to see has been added here.)}}%
%    \label{Scalar_potential}%
%\end{figure}

\section{Consistency check on the UV cutoff}

To study the scale at which nonrenormalizable interactions appear we follow the method used in \cite{JP1,JP2,JP3,JP4}}. We analyze nonrenormalizable terms that come from either the new-FI D term or the new GM F term. We can generalize the strategy introduced in Ref.~\cite{JP4} in the following way. Let us write the Lagrangian 
density of the new FI and GM terms as
\begin{eqnarray}
\mathcal{L} = {1\over 2} D^2 - \Xi D+ \sum_{n\geq 0, m\geq 0} D^{-n} M_{Pl}^{-m} O^{2n+m +4}_{\textrm{newFI}} + \mathcal{L}_{GM}
\end{eqnarray}
where $\Xi$ is the new-FI-term constant and $O^{2n+m +4}_{\textrm{newFI}}$ are nonrenormalizable local operators 
of scaling dimension $2n+m+4$. All terms with $m>0$ vanish in the global supersymmetry limit $M_{pl} \rightarrow \infty$ and $D\sim \Xi$ fixed. This shows that the strongest constraint on the UV cutoff comes either from the term with $m=0$ or from $\mathcal{L}_{GM}$. The $m=0$ terms gives
\begin{eqnarray}
  \Lambda_{cut}^2 \lesssim D \sim \Xi \sim HM_{pl} \sim M_S^2 \implies \Lambda_{cut} \lesssim \sqrt{HM_{pl}} ,\label{NewFI_cutoff_con};
\end{eqnarray}
this is the same bound obtained in~\cite{JP4}. Moreover, in the limit of $M_{pl} \rightarrow \infty$ and in the 
superconformal gauge ($s_0 = \bar{s}_0 = M_{pl}e^{K/6M_{pl}^2}$),  the $\mathcal{L}_{GM}$ term in the Lagrangian 
reduces to 
\begin{eqnarray}
&& \mathcal{L}_{\textrm{GM}}
\nonumber\\
&& =  -\frac{1}{W}(\bar{\lambda}_U P_L\lambda_U)(D_V^2 - \hat{F}_V^-\cdot\hat{F}_V^- - 2  \bar{\lambda}_VP_L\cancel{\mathcal{D}}\lambda_V)
 \nonumber\\
 &&-\frac{1}{W}(\bar{\lambda}_V P_L\lambda_V)(D_U^2 - \hat{F}_U^-\cdot\hat{F}_U^- - 2  \bar{\lambda}_UP_L\cancel{\mathcal{D}}\lambda_U)
 \nonumber\\
 && -2\frac{1}{W}\bar{\lambda}_U(-\frac{1}{2}\gamma\cdot \hat{F}_U + iD_U)  P_L(-\frac{1}{2}\gamma\cdot \hat{F}_V + iD_V)\lambda_V
 \nonumber\\
 &&  -\Big(\frac{W_iF^i}{W^2}\Big)(\bar{\lambda}_U P_L\lambda_U) (\bar{\lambda}_V P_L\lambda_V)\nonumber\\
 && +\Big(\frac{W_{ij}}{2W^2}+\frac{W_iW_j}{W^3}\Big)\bar{\chi}^iP_L\chi^j(\bar{\lambda}_U P_L\lambda_U) (\bar{\lambda}_V P_L\lambda_V)
\nonumber\\
&&+ \Big(\frac{W_i}{W^2}\bar{\chi}^iP_L\Big)(\bar{\lambda}_U P_L\lambda_U) \sqrt{2}P_L(-\frac{1}{2}\gamma\cdot \hat{F}_V + iD_V)\lambda_V
\nonumber\\
&& +\Big(\frac{W_i}{W^2}\bar{\chi}^iP_L\Big) (\bar{\lambda}_V P_L\lambda_V)\sqrt{2}P_L(-\frac{1}{2}\gamma\cdot \hat{F}_U + iD_U)\lambda_U .
\end{eqnarray}
This equation includes both renormalizable and nonrenormalizable terms. From the latter we find new constraints on the cutoff scale $\Lambda_{cut}$. They are found to be
\begin{eqnarray}
&& \textcircled{\small{1}} ~ \frac{1}{W} \lesssim \Lambda_{cut}^{-3},  ~\textcircled{\small{2}} ~\frac{D_V}{W} \lesssim \Lambda_{cut}^{-1} , ~ \textcircled{\small{3}} ~ \frac{W_iF^i}{W^2} \lesssim \Lambda_{cut}^{-2}, \nonumber\\
 &&  \textcircled{\small{4}}~ \Big(\frac{W_{ij}}{2W^2}+\frac{W_iW_j}{W^3}\Big) \lesssim \Lambda_{cut}^{-5}, ~ \textcircled{\small{5}}~ \frac{W_i}{W^2} \lesssim \Lambda_{cut}^{-4}, ~ \textcircled{\small{6}}~ \frac{W_iD_V}{W^2} \lesssim \Lambda_{cut}^{-2}. \label{constraints_GM}
\end{eqnarray}
Let us examine each of them separately. The constraint \textcircled{\small{1}} reduces to 
\begin{eqnarray}
  \textcircled{\small{1}} ~ \Lambda_{cut} \lesssim W^{1/3}. \label{Con1} 
\end{eqnarray}
From \textcircled{\small{1}} and \textcircled{\small{2}}, we find
\begin{eqnarray}
    \textcircled{\small{2}} ~ D_V \lesssim \Lambda_{cut}^2 .\label{Con2}
\end{eqnarray}
From \textcircled{\small{5}} and \textcircled{\small{3}}, we find
\begin{eqnarray}
    \textcircled{\small{3}} ~ F^i \lesssim \Lambda_{cut}^2.\label{Con3}
\end{eqnarray}
From \textcircled{\small{1}} and \textcircled{\small{4}}, we find
\begin{eqnarray}
  \textcircled{\small{4}}~  \frac{W_{ij}}{2}+\frac{W_iW_j}{W} \lesssim \Lambda_{cut} \implies \frac{W_{ij}}{2} \lesssim \Lambda_{cut} \quad \mathrm{and} \quad  \frac{W_iW_j}{W} \lesssim \Lambda_{cut}.\label{Con4}
\end{eqnarray}
From \textcircled{\small{1}} and \textcircled{\small{5}}, we find
\begin{eqnarray}
 \textcircled{\small{5}}~   W_i \lesssim \Lambda_{cut}^2.\label{Con5}
\end{eqnarray}
Using \textcircled{\small{5}} and \textcircled{\small{6}}, we find that the contraint \textcircled{\small{6}} in Eq.~\eqref{constraints_GM} reduces to the constraint \textcircled{\small{2}} in~\eqref{Con2}. Notice that Eqs.~\eqref{NewFI_cutoff_con} and~\eqref{Con2} bound the value of the cutoff scale as follows
\begin{eqnarray}
    D_V \sim \Lambda_{cut}^2 \sim M_S^2 \sim HM_{pl} \implies \Lambda_{cut} \sim M_S  \sim \sqrt{HM_{pl}}.
\end{eqnarray}
This is essential for the bounds coming from the new FI term and the gaugino mass to be compatible with each other. With this result, the constraints \textcircled{\small{1}}, \textcircled{\small{3}}, \textcircled{\small{4}}, and \textcircled{\small{5}} show that 
\begin{eqnarray}
  M_S^3 \lesssim W, \quad F^i \lesssim M_S^2, \quad W_{ij} \lesssim M_S, \quad W_i \lesssim M_S^2, \label{final_constraints}
\end{eqnarray}
or equivalently
\begin{eqnarray}
  HM_{pl} \sqrt{HM_{pl}} \lesssim W, \quad F^i \lesssim HM_{pl}, \quad W_{ij} \lesssim \sqrt{HM_{pl}}, \quad W_i \lesssim HM_{pl},
\end{eqnarray}
which can be satisfied for the values $F^i\lesssim HM_{pl}$ and $W \lesssim \mathcal{O}(HM_{pl}^2) $ that we used
in our supergravity model of inflation with the new FI terms. Summing all up we find that the constraints in 
Eqs.~\eqref{Con1} through Eq. \eqref{Con5}, which arise from the gaugino mass Lagrangian \eqref{GM}, can be satisfied
 while also satisfying Eq.~\eqref{NewFI_cutoff_con}.

Let's study in detail the constraints following from Eq.~\eqref{final_constraints}. We begin with looking at the first one by writing the superpotential considered in Ref.~\cite{JP4} 
\begin{eqnarray}
    W(T,z^i) = \underbrace{W_0 + Ae^{-aT}}_{\textrm{hidden sector}} + \underbrace{W_{MSSM}(z^i)}_{\textrm{observable sector}},
\end{eqnarray}
where $W_0 = -cA$ and $c = (1+a/3)e^{-a/2}$. From Ref.~\cite{JP4}, we see that $M_S^4 = \Lambda + \frac{a^2A^2e^{-a}}{3} \approx (\frac{aAe^{-a/2}}{\sqrt{3}})^2$. At the minimum $(T = 1/2, z \sim 0)$, the value of the superpotential is $|W|_{vac} = \frac{a}{3}e^{-a/2}A = M_S^2/\sqrt{3}$. Therefore, the constraint $M_S^3 \lesssim W$ is satisfied at the 
post-inflation vacuum as long as $M_S < M_{pl} = 1$. For large $T$ along the inflationary plateau, the superpotential
 reduces to $|W_0| = Ae^{-a/2} +aAe^{-a/2}/3= \frac{\sqrt{3}M_S^2}{a}+M_S^2/\sqrt{3} $, which automatically satisfies the constraint due to the previous condition for the vacuum.
The second condition in~\eqref{final_constraints} holds because the scalar part of $F^i$ in Eq.~\eqref{matter_aux} is given by $F^i \sim - 3e^{K/2}G^{i\bar{j}}(\bar{W}_{\bar{j}}+K_{\bar{j}}\bar{W}) \sim 0 $ along the inflationary trajectory (i.e. $z^i \approx 0$). The third and fourth ones can be easily satisfied due to the fact that the derivatives of the superpotential with respect to the matter scalars are either of order of a low scale, much smaller than the supersymmetry breaking one, or vanish along the inflationary trajectory.

\section{Conclusion}

 We presented a simple, predictive model with moduli stabilization without backreaction on the scalar field dynamics and  {\it parametrically large} gaugino masses. We saw that the gaugino masses in our model can be given a value close to the
 Hubble scale
\begin{eqnarray}
    M_{\textrm{gaugino}} = \frac{\left<D_V\right>^2}{m_{3/2}M_{pl}^2} = \frac{M_S^4}{m_{3/2}M_{pl}^2} \sim \mathcal{O}(H)=\mathcal{O}(10^{-5}M_{pl}),\label{GM_with_mass_dim}
\end{eqnarray}
where $\left<D_V\right>= \xi = M_S^2 = HM_{pl}$, $m_{3/2} \sim H$, and $M_S$ is the SUSY-breaking mass scale, 
which is of order $10^{-2.5}M_{pl}$. This means that gauginos can be integrated out at energies below the Hubble scale, 
and thus are to all effects undetectable. Note that the gaugino masses \eqref{GM_with_mass_dim} do not depend on the 
VEVs of matter fields --which either vanish  at the post-inflation vacua or give negligibly small contributions to the masses.

As a further study, it would be interesting to seek string realization of the gaugino mass term \eqref{GM} proposed here. The hint for this may come from the four-dimensional gaugino-mass term in Ref.~\cite{Grana:2020hyu}, which is obtained by dimensional reduction of a fermionic space-filling D$p$-brane action carrying both RR and NSNS fluxes.

\subsection*{Acknowledgments} 
H.J. is supported in part by Incheon National University RIBS Grant in 2023, and in part by Basic Science Research Program through the National Research Foundation of Korea (NRF) funded by the Ministry of Education through the Center for Quantum Spacetime (CQUeST) of Sogang University (NRF-2023R1A2C2005360). H.J. would like to thank YITP at Kyoto University for its kind hospitality. M.P. is supported in part by NSF grant PHY-2210349 and by the Leverhulme Trust through a Leverhulme Visiting Professorship at Imperial College, London.

\end{document}